\documentstyle[11pt]{article}
\begin{document}
\vspace{20mm}

\title{Solitons on Noncommutative Torus as Elliptic Calogero Gaudin Models,
 Branes and Laughling wave functions}
\author{Bo-Yu Hou$$ \thanks{Email:byhou@phy.nwu.edu.cn}, \hspace{5mm}
       Dan-Tao Peng$$ \thanks{Email:dtpeng@phy.nwu.edu.cn}, \hspace{5mm}
       Kang-Jie Shi$$ \thanks{Email:kjshi@phy.nwu.edu.cn}, \hspace{5mm}
       Rui-Hong Yue$$ \thanks{Email:yue@phy.nwu.edu.cn}\\[3mm]
       Institute of Modern Physics, Northwest University\\
       Xi'an, 710069, China}
\date{}
\maketitle

\begin{abstract}
For the noncommutative torus ${\cal T}$, in case of the N.C. parameter
$\theta = \frac{Z}{n}$, we construct the basis of Hilbert space ${\cal
H}_n$ in terms of $\theta$ functions of the positions $z_i$ of $n$
solitons. The wrapping around the torus generates the algebra ${\cal
A}_n$, which is the $Z_n \times Z_n$ Heisenberg group on $\theta$
functions. We find the generators $g$ of an local elliptic $su(n)$, which
transform covariantly by the global gauge transformation of ${\cal A}_n$.
By acting on ${\cal H}_n$ we establish the isomorphism of ${\cal A}_n$ and
$g$. We embed this $g$ into the $L$-matrix of the elliptic Gaudin and C.M.
models to give the dynamics. The moment map of this twisted cotangent
$su_n({\cal T})$ bundle is matched to the $D$-equation with
Fayet-Illiopoulos source term, so the dynamics of the N.C. solitons
becomes that of the brane. The geometric configuration $(k, u)$ of the
spectral curve ${\rm det}|L(u) - k| = 0$ describes the brane
configuration, with the dynamical variables $z_i$ of N.C. solitons as the
moduli $T^{\otimes n} / S_n$. Furthermore, in the N.C. Chern-Simons
theory for the quantum Hall effect, the constrain equation with
quasiparticle source is identified also with the moment map eqaution of
the N.C. $su_n({\cal T})$ cotangent bundle with marked points. The
eigenfunction of the Gaudin differential $L$-operators as the Laughlin
wavefunction is solved by Bethe ansatz.

\vspace{.5cm}
\end{abstract}

\setcounter{equation}{0}

\section{Introduction}

\indent

The development of soliton theory on noncommutative geometry is rather
impressive (details of reference e.g. see review \cite{H, DN}). The paper
\cite{GMS} gives the solitons expressed by the projection operators.
Witten \cite{W} gives the partial isometry operators. The paper \cite{HKL}
uses the partial isometry operators to generate the solitons on
noncommutative plane.

Recently, the solitons on the N.C. torus has attracted a lot of interest
\cite{BKMT, SS, MM, B, GHS, TKMS, KMT, Bars}. On the compactified torus
\cite{CDS}, the duality, the Morita equivalence and the orbifolding have
been studied \cite{SW}.

The equivalence class of projection operator on torus is given by
\cite{R} in terms of $U_1$ and $U_2$. The noncommutative algebra ${\cal
A}$ generated by $U_i$ ($U_1 U_2 = U_2 U_1 e^{i\theta}$) wrapping the
torus has been given in \cite{CDS, SW, KS} in terms of the matrix
difference operator. The matrix acts on a $U(n)$ bundle $V_n$ with trivial
connection while the difference acts on a $U(1)$ bundle ${\cal L}$. The
Hilbert space ${\cal H} = V_n \otimes {\cal L}$ acted  by ${\cal A}$ is
given by the vector functions of {\it real} variables. The local covariant
derivative operator $\bigtriangledown_i$ acted on ${\cal L}$ is given
also. Their commutator gives the constant curvature related to $\theta$.
Obviously the double periodicity of wave functions on torus can not be
given explicitly by real variable functions. The Schwarz space of smooth
functions $S({\bf R})$ of rapid decrease real functions provides a
bimodule between ${\cal A}_\theta$ and ${\cal A}_{\frac{1}{\theta}}$ under
Morita equivalence \cite{MM, KMT}. Starting from the Gaussian function
$e^{i\pi y^2/\theta}$ as the Schwarz function, Boca \cite{B} constructed
the projection operators in terms of $\theta$ functions. This projection
operator satisfies the BPS like selfduality condition \cite{BKMT, DKL},
thus gives constant curvature. But Boca obtained the explicit expressions
only for the case with modulus equals $\frac{1}{Z}$, moreover, it is in
terms of the products of two $\theta$ functions depends seperately on
$U_1$ and $U_2$, so the symplectic structure of noncommutative torus is
unclear. Gopakumar et al \cite{GHS} starting from the same Gaussian
function but orbifolding the double periodic multisoliton solution on the
entire N.C. ${\bf R}^2$ into a single soliton solution on the torus. They
employed the $p q$ representation \cite{BCZ, Zak} on a dual lattices for
conjugate variables $p$ and $q$, which provides a basis of simultaneous
eigenstates of commutative $U_1$ and $U_2$ with $\theta = 2 \pi A$, where
$A$ is an integer. Thus they succeeded in constructing the soliton on the
so-called integral torus with double periodic wave functions. The
noncommutative symplectic complex structure on torus appears explicitly
and the corresponding Weyl-Moyal transformation is realized as double
series of $U_1^{\frac{1}{A}}$ and $U_2^{\frac{1}{A}}$. But since the loop
$U_1$ and $U_2$ become commutative in this degenerated $\theta = 2 \pi A
(e^{i \theta} = 1)$ case, they obtained just a unique projection operator,
corresponding to only one soliton in each lattice (i.e. in the torus).

Now in this paper, by $n$ times orbifolding such integral torus ${\cal T}$
into a ${\cal T}_n$, we construct the $n$-th multi-soliton solution with
$n$ zeros $z_i$ in ${\cal T}$, such that the moduli space of this kind of
multiple soliton becomes ${\cal T}^{\otimes n} / {S_n}$. Meanwhile, the
fundamental cycles $U_i$ wrapping around ${\cal T}$, will be subdivided
$n$ times into corresponding $W_i$ upon ${\cal T}_n$, and $W_1 W_2
W_1^{-1} W_2^{-1} = e^{i \frac{2 \pi}{n}}$. The torus ${\cal T}$ now
covers ${\cal T}_n$ $n$ times, i.e. the original $2$-brane on ${\cal T}$
turns to be $n$ $2$-branes with a $U(n)$ symmetry $G \sim U(n)$. The {\it
global} $W_i$ generates an ${\cal A}_n$ algebra as the $Z_n \times Z_n$
Heisenberg group $G_{\cal H}$. Moreover, using the modules $z_i$,
$\bar{z}_i (\sim \partial_i)$, we construct the {\it local} operators
$E_\alpha$ which generates the algebra $g \sim u_n({\cal T})$ of $G$.

Then for the cotangent bundle of this $su_n({\cal T})$ bundle on torus $u$
twisted by $U_i$, by using a moment map with sources at the marked points,
we obtain the quantum elliptic Gaudin Lax differential operator $L(u)$ as
its $n \times n$ covariant Hermitean section. In case of only one marked
point we gauge transform this $L(u)$ into the elliptic Calogero-Moser Lax
operator. Obviously the moduli space becomes symplectic ($z_i, \bar{z}_i
\rightarrow q_i, p_i$) with Hamiltonian flows generated by C.M.
Hamiltonian. The spectral curve determines the "shape" in target space of
the brane configuration in DHWW construction. The moment map equation
serves as the D-equation. The dynamics of brane is given by the evolution
of the $n$ solutions $z_i(t)$ of the C.M. equation.

Furthermore, we consider this moment map equation as the Gaussian
constrain equation in QHE. The quantum eigenfunctions of Hamiltonian
operators of Gaudin models becomes the Langhlin wave functions on torus.

In the next section, after review the construction by Gopakumar et. al
\cite{GHS} for the soliton on the integral torus ${\cal T}$, we show by $n
\times n$ times orbifolding this ${\cal T}$ into ${\cal T}_n$, that the
one dimensional trivial Heisenberg group i.e. the double periodicity under
wrapping by $U_i$ on ${\cal T}$ in paper \cite{GHS} will be refined into a
$Z_n \times Z_n$ $G_{\cal H}(n)$ generated by loops $W_i (\equiv
U_i^\frac{1}{n})$ around ${\cal T}_n$.

In section 3, we use the $\theta$ functions of the central "positions"
$z_i$ of multiple solitons to give an explicit expression for the basis of
the Hilbert space ${\cal H}_n$. The ${\cal A}_n$ algebra turns to be
the $Z_n \times Z_n$ Heisenberg group $G_{\cal H}(n)$ acting on $\theta$
functions \cite{Tata}. The local infinitesimal translation operator
$E_\alpha$ is realized as the derivative operator generating the
$su_n({\cal T})$ algebra on torus. The $E_\alpha$ behaves covariantly
under gauge transformation by $G_{\cal H}(n)$. We establish the
isomorphism of the algebra ${\cal A}_n$ and the $su_n({\cal T})$ by
applying them on ${\cal H}_n$, they both become identical to the $G_{\cal
H}(n)$.

Later, in section 4, we embed this $su_n({\cal T})$ derivative operator as
the "quantum" operators in the representation of the transfer matrix (Lax
operator) $L(u)$ of the elliptic Gaudin model i.e. $L(u)$ is obtained
from the classical Yang-Baxter $r$-matrix $r(u) = \sum_\alpha w_\alpha(u)
I_\alpha \otimes I_\alpha$ acted on $V_n \otimes V_n$ space by replacing
one $V_n$ into a quantum space, i.e. replacing the Heisenberg matrix
$I_\alpha$ on it by the $su_n({\cal T})$ operator $E_\alpha$. The
$r$-matrix is nondynamic and depends only on the spectral (evalution)
parameter $u$ through the meromorphic sections $w_\alpha(u)$. The double
Heisenberg properties of $w_\alpha(u)$ and hence of $r(u)$ ensures $L(u)$
to be a section on a twisted $su(n)$ bundle. Then we gauge transform the
Gaudin Lax into the Lax of the elliptic Calogero-Moser models. This Lax is
equivalent to the elliptic Dunkle operators representing Weyl reflections.
The trace of the quadratic of $L$ gives the Hamiltonian of the C.M. which
is assumed to give dynamics of the N.C. solitons on the brane. As in
section 3, we have shown that the isomorphism of the N.C. {\it $z_i$ and
$\bar{z}_i$} to the {\it $z_i$ and local translation ${\partial_i}$} in
N.C. $R^2$ case; under orbifold $\frac{R^2}{Z \otimes Z}$, becomes the
isomorphism of the {\it $Z_n \times Z_n$ Heisenberg $W_\alpha$} in ${\cal
A}_n$ to {\it the $E_\alpha$} in $su_n({\cal T})$. So the correspondence
of noncommutative $\partial_i (\sim {\it \bar{z}_i})$ and ${\it z_i}$ in
$E_\alpha$ to dynamical ${\it p_i}$ and ${\it q_i}$ in $L(u)$, naturally
endows the N.C. torus with symplectic structure. Thus naturally the
solitons on brane satisfy the dynamics of Calogero Gaudin systems.

In section 5 we describe the brane picture of the N.C. solitons. The
moment map for reducing this twisted cotangent $su_n({\cal T})$ bundle is
matched to the $D$-equations with Fayet-Illiopoulos source term, so the
dynamics of the N.C. solitons becomes that of the brane with impurities.
The geometric configuration of the spectral curve ${\rm det}|L(u) - k| =
0$ describes the brane configuration $K(u) = 0$, with the dynamical
variables $z_i$ of N.C. solitons as the modules describing the brane.

Furthermore, in section 6 we find the Laughlin wavefunction for quantum
Hall effect (QHE) on N.C. torus. In the N.C. Chern-Simons theory for the
QHE, the constrain equation with quasiparticle source is identified also
with the moment map equation of the N.C. $su_n({\cal T})$ cotangent bundle
with marked points. In the following we show that the Laughlin wave
function is given as a special case of the Bethe ansatz solution of Gaudin
differencial $L$ oprators. In the B.A. given by Felder and Varchenko
\cite{FV4}, the so called "spectral parameter $\zeta$ is generic. We
specialize it to be matched with the physical quantities in QHE etc. in
the following way: we analyze succeedingly how it is related to the
twisting phase of the B.A. function, to the twisted boundary condition and
boundary Hamiltonian. Then we consider the algebraic geometrical form of
the eigenfunction of C.M., so we find that this phase shift is determined
by the filling fraction in the Gaussian constrain in quantum Hall fluid
equation. Thus we confirm  that this specialized B.A. eigenfunction is
really the Laughlin wave function on torus. We give the solution for the
case of only one quasiparticle, i.e. the Gaudin bundle with one marked
point, which is equivalent to the C.M. bundle. It is easy to generalize it
to the multi-quasiparticle case, i.e. Gaudin bundle with more marked
points.

In the last section, we shortly describe the subjects which will be
investigated later.

\section{Solitons on the "integral torus" \cite{GHS} and its further
orbifolding}

\indent

In this section we at first shortly review the result of paper \cite{GHS}
for solitons on nocommutative space ${\bf R}^2$: $[\hat{x}_1, \hat{x}_2]
= i \Theta$. This ${\bf R}^2$ has been orbifolded to a torus ${\cal T}$
with periodicities $L$ and $\tau L$.

The generators of the fundamental group of ${\cal T}$ are:
\begin{equation}
U_1 = e^{- i \hat{y}^2 l}, \hspace{10mm} U_2 = e^{i l (\tau_2 \hat{y}^1 -
\tau_1 \hat{y}^2)},
\end{equation}
where the normalized length $l \equiv L \Theta^{\frac{1}{2}}$, $\hat{y}^i
= \frac{1}{\sqrt{\Theta}}\hat{x}^i$, $(i = 1, 2)$ and
\begin{equation}
U_1 U_2 = U_2 U_1 e^{- i\tau_2 l^2}
\end{equation}
As in \cite{GHS}, let us consider the case $\frac{\tau_2 l^2}{2\pi} \in
{\bf N}$ (or ${\bf Z}_+$) i.e. the normalized area $A = \frac{\tau_2
l^2}{2 \pi} =$ the $B$ flux passing through the torus, is an integer. The
projection operators constructed by \cite{GHS} has its image spanned by
the lattice of coherent states
\begin{equation}
U_1^{j_1} U_2^{j_2}|0\rangle, (j_1, j_2 \in {\bf Z}^2)
\end{equation}
here $|0\rangle$ is the Fock Bargmann vacuum vector: $a|0\rangle = 0$,
where $a = \frac{1}{\sqrt{2}}(\hat{y}^1 + i\hat{y}^2)$, $a^\dagger =
\frac{1}{\sqrt{2}}(\hat{y}^1 - i\hat{y}^2)$, $[a, a^\dagger] = 1$. They
found a particular linear combination:
\begin{equation}
\label{wavefunction}
|\psi\rangle = \sum_{j_1, j_2}c_{j_1, j_2}U_1^{j_1}U_2^{j_2}|0\rangle
\end{equation}
that satisfies
\begin{equation}
\label{orthonormal}
\langle\psi|U_1^{j_1}U_2^{j_2}|\psi\rangle = \delta_{j_1 0}\delta_{j_2 0}.
\end{equation}
Then they found the projection operator
\begin{eqnarray}
P & = & \sum_{j_1, j_2} U_1^{j_1} U_2^{j_2} |\psi\rangle
\langle\psi|U_2^{- j_2} U_1^{- j_1} \nonumber\\
P^2 = P.
\end{eqnarray}
Since the orthonormalities (\ref{orthonormal}), we have $U_i P U_i^{-1} =
P$, $(i = 1, 2)$, i.e. $P$ actually acts on ${\cal T}={\cal R}^2 / {\bf
Z} \times {\bf Z}$ i.e. with periods $U_i$ on $R^2$.

To find the solution satisfying the orthonormality condition
(\ref{orthonormal}), the wave function in (\ref{wavefunction}) on $R^2$
has been found \cite{GHS} in terms of the $p q$ representation \cite{BCZ,
Zak}:
\begin{equation}
\label{pq}
|p q \rangle =
\sqrt{\frac{l}{2\pi}}e^{-i\tau_1(\hat{y}^2)^2/{2\tau_2}} \sum_j e^{ijlp}
|q + jl\rangle.
\end{equation}
where $| q \rangle$ is an eigenstate of $\hat{y}^1$: $\hat{y}^1 |q\rangle
= q |q\rangle$, such that
\begin{equation}
e^{i l \tau_2 \hat{y}^1}|q\rangle = e^{ilq\tau_2}|q\rangle, \hspace{5mm}
e^{- i l \hat{y}^2}|q\rangle = |q + l\rangle.
\end{equation}
Obviously, in case $A$ being an integer, $| p q \rangle$ is the common
eigenstate of $U_1$ and $U_2$:
\begin{equation}
U_1 | p q \rangle = e^{-i l p} | p q\rangle, \hspace{1cm} U_2 | p q
\rangle = e^{i l \tau_2 q} | p q \rangle.
\end{equation}
Remarks: From the completeness of the coherent states \cite{Perelomor}, we
know that $| p q \rangle$ is defined in the double periods: $0 \leq p \leq
\frac{2\pi}{l}, 0\leq q\leq l$, and constitute an orthonormal and complete
basis for the Hilbert space ${\cal H}_A = {\cal H}_{{\bf R}^2}/A$, if $A
\geq 2$. In case of $A = 1$, if we choose just one coherent state for any
unit cell, then all the state excluding the vacuum state maybe taken as a
complete and minimal system of states. Bars \cite{Bars} recently has
discussed the solutions with size smaller than this extremal one.

By using $|q\rangle = \exp(\frac{(a^\dagger)^2}{2} + a^\dagger q +
\frac{q^2}{2})|0\rangle$, we find the sum in wave function (\ref{pq})
equals the $\theta$ function
\begin{equation}
\label{pq-wave}
\langle p q | 0 \rangle = \sqrt{\frac{l}{2 \pi}} e^{\frac{\tau}{2 i
\tau_2}q^2}\theta(\frac{(p + i q)A}{\tau l}, \frac{A}{\tau}),
\end{equation}
where $\tau = \tau_1 + i \tau_2$. Then by modular transformation of
(\ref{pq-wave}) or directly from (\ref{pq}) by using the Poisson
resummation formula we obtain as in \cite{GHS}:
\begin{equation}
\label{wave-func}
C_0(p, q) \equiv \langle p q | 0 \rangle =
\frac{1}{\pi^{\frac{1}{4}}\sqrt{l}}\exp(-\frac{\tau}{2i\tau_2}p^2 + i p q)
{\cal \theta}_{00}(\frac{q + p \tau / \tau_2}{l}, \frac{\tau}{A}).
\end{equation}
The orthonormality condition (\ref{orthonormal}) becomes
\begin{eqnarray}
\delta_{j_1, 0}\delta_{j_2, 0} & = & \int_0^{\frac{2 \pi}{l}} dp \int_0^l
dq e^{- i j_1 l p + i j_2 l \tau_2 q}|\Psi(p, q)|^2\nonumber\\
& = & \int_0^{\frac{2 \pi}{l}} dp \int_0^{\frac{l}{A}} dq e^{- i j_1 l p +
i j_2 l \tau_2 q}|\tilde{c}(p, q)|^2 \sum_{n = 0}^{A - 1}\left | C_0(p, q
+ n \frac{l}{A})\right |^2.
\end{eqnarray}
Thus, the paper \cite{GHS} found that
\begin{equation}
\label{Psi}
\langle p q | \psi \rangle \equiv \Psi(p, q) = \tilde{c}(p, q) C_0(p, q) =
\frac{C_0(p, q)}{\sqrt{2 \pi / A \sum_{n = 0}^{A - 1}|C_0(p, q + n
l/A)|^2}}.
\end{equation}

Now we turn to the further orbifolding. Let $W_1 = U_1^{\frac{1}{n}}$,
then
\begin{equation}
C_1(p, q) \equiv \langle p q | W_1 | 0 \rangle = \langle p, q +
\frac{l}{n} | 0 \rangle = \frac{1}{\pi^{\frac{1}{4}} \sqrt{l}} \exp(-
\frac{\tau p^2}{2 i \tau_2} + i p (q + \frac{l}{n})) \theta_{0,
\frac{1}{n}}(\frac{q + \frac{p \tau}{\tau_2}}{l}, \frac{\tau}{A}),
\end{equation}
here we have chosen $A$ and $n$ relatively prime.

Similarly
\begin{eqnarray}
C_\beta(p, q) \equiv \langle p q | W_1^\beta | 0 \rangle & = &
\frac{1}{\pi^{\frac{1}{4}}\sqrt{l}} \exp(- \frac{\tau p^2}{2 i \tau_2} + i
p (q + \frac{\beta}{n} l)) \theta_{0, 0}(\frac{q +
\frac{p\tau_1}{\tau_2}}{l} + \frac{\beta}{n}, \frac{\tau}{A})\nonumber\\
& \equiv & \frac{1}{\pi^{\frac{1}{4}}\sqrt{l}} \exp(- \frac{\tau p^2}{2 i
\tau_2} + i p (q + \frac{\beta}{n} l)) \theta_{0, \frac{\beta}{n}}(z,
\frac{\tau}{A}),
\end{eqnarray}
here $z \equiv y_1 + i y_2$, the map $p, q$ to $z$ is given as in
\cite{GHS} by the Weyl Moyal transformation. The effect of $W_1$ is to
shift $z$ to $z + \frac{1}{n}$, or equivalently shift the characteristic
$\beta$ of $\theta_{\alpha, \beta}$ to $\beta + \frac{1}{n}$.

Constructing
\begin{equation}
\label{Psi_beta}
\langle p q |\Psi_\beta\rangle = \frac{C_\beta(p, q)}{\sqrt{\frac{2\pi}{A}
\sum_{n=0}^{A-1}|C_\beta(p, q + n \frac{l}{A})|^2}},
\end{equation}
then $|\Psi_{\beta + n}\rangle = e^{-i l p}|\Psi_\beta\rangle$ and
$|\Psi_\alpha \rangle$ ($\alpha = 0, 1, \cdots, n-1$) are linearly
independent.

Let
\begin{equation}
\label{P_beta}
P_\beta = \sum_{j_1, j_2} U_1^{j_1} U_2^{j_2} | \Psi_\beta \rangle \langle
\Psi_\beta | U_2^{- j_2} U_1^{- j_1},
\end{equation}
then
\begin{equation}
P_{\beta}^2 = P_{\beta}
\end{equation}
\begin{equation}
W_1^{\beta_1} P_{\beta_2} W_1^{- \beta_1} = P_{\beta_1 + \beta_2}.
\end{equation}

As in paper \cite{SW}, We may also orbifold the torus along the
$\overrightarrow{\tau}$ direction $n$ times. Let $W_2 =
U_2^{\frac{1}{n}}$, then
\begin{eqnarray}
C_{\alpha}^\prime(p, q) & \equiv & \langle p q | W_2^{\alpha}| 0 \rangle =
e^{i \frac{\alpha}{n} l \tau_2 q} \langle p + \frac{\tau_2 l}{n}\alpha,
q | 0 \rangle \nonumber\\
& = & \frac{1}{\pi^{\frac{1}{4}}\sqrt{l}}\exp(- \frac{\tau p^2}{2 i
\tau_2} + i (p + \frac{\alpha}{n} l \tau_2) q) \theta_{\frac{\alpha}{n} A,
0}(\frac{q + \frac{p \tau}{\tau_2}}{l}, \frac{\tau}{A}).
\end{eqnarray}
Obviously, the $W_2$ shifts $z$ to $z + \frac{\tau}{n}$, i.e. shift the
$\alpha$ of $\theta_{\alpha, \beta}$.

Subsequently, we may construct $|\psi_{\alpha}^\prime \rangle$ and
$P_{\alpha}^\prime$ as in (\ref{Psi_beta}) and (\ref{P_beta}). But the set
$P_{\alpha}^\prime$, $(\alpha = 1, 2, \cdots, n)$ are not independent from
the set $P_\beta$, $(\beta = 1, 2, \cdots, n)$, and in either basis the
$W_1$ or(and) $W_2$ matrix are not constant. Actually, since the target
space ${\cal H}_{\cal T}$ of $P$ on total torus ${\cal T}$ has been
subdivided into ${\cal H}_n$ on torus ${\cal T}_n$ as described by
\cite{SW}. One may find a basis of ${\cal H}_n$ such that $W_1$ and $W_2$
become the $n \times n$ irreducible matrix representation of the
Heisenberg group
\begin{equation}
W_1 W_2 = W_2 W_1 \omega, \hspace{1cm} \omega^n = 1,
\hspace{1cm} W_1^n = W_2^n = 1.
\end{equation}

In next section we will construct explicitly this basis in terms of
$\theta_{\alpha \beta}$ functions. But before that, let us compare with
the case of N.C. plane, where the generic $n$ solitons solution is
$\prod_{i = 1}^n(a^\dagger - z_i)|0\rangle$, with $n$ soliton centers at
$z_i$. Similarly, we will introduce the $z_i$ for the location of the
centers of $n$-solitons solutions on the torus ${\cal T}$, such that the
moduli space of the $n$-soliton solution is ${\cal T}^{\otimes n}/S_n$.

\section{Noncommutative algebra ${\cal A}_n$, Hilbert space ${\cal H}_n$,
Heisenberg group $Z_n \times Z_n$ and $su(n)$ algebra, in case $[U_1, U_2]
= 0$}

\subsection{The Hilbert space}

\indent

As in \cite{SW}, usually the Hilbert space ${\cal H}_{\cal T}$ can be
written as the direct product of a $su(n)$ trivial bundle $V_n$ and an
$U(1)$ line bundle ${\cal L}$: ${\cal H}_{\cal T} = V_n \otimes {\cal L}$.
On the $V_n$ acts the $W_1$ and $W_2$ matrices, on the ${\cal L}$ acts the
covariant defference operators $\hat{\bf V}_i$ (the notation as that in
\cite{CDS}), and $W_i \otimes \hat{\bf V}_i = U_i$. Thus, the operators
$U_i$ are matrix difference operators acting on vector functions $v_a, (a
= 1, 2, \cdots, n)$. But now it happens that in case of commutative $U_1$
and $U_2$, this space $V_n$ is identical to the whole ${\cal H}_n$ on
subtorus ${\cal T}_n$ in the following way.

The basis vectors of the {\it Hilbert space ${\cal H}_n$} are
\begin{eqnarray}
\label{basis}
v_a & = & \sum_{b=1}^n F_{-a, b}, (a= 1, 2, \cdots, n), \nonumber\\
F_{\alpha} & \equiv & F_{\alpha_1, \alpha_2} = e^{i\pi n \alpha_2}
\prod_{j=1}^n \sigma_{\alpha_1, \alpha_2}(z_j - \frac{1}{n}\sum_{k=1}^n
z_k),
\end{eqnarray}
here $\alpha \equiv (\alpha_1, \alpha_2) \in Z_n \times Z_n $, and
$$
\sigma_{\alpha}(z) = \theta \left [
\begin{array}{c}
\frac{1}{2} + \frac{\alpha_1}{n}\\
\frac{1}{2} + \frac{\alpha_2}{n}\\
\end{array}\right ](z, \tau).
$$
(In the following, we use the modulus $\tau$ to represent the
$\frac{\tau}{A}$ in section 2).

Remark: The $\Psi$ (\ref{Psi}) has been normalized, but it is not entire,
i.e. it has poles and is nonanalytic. Our basis $v_a$ is entire, they have
$n$ zeros in a torus, i.e. they span the space of weight $n$ quasiperiodic
functions (sections) \cite{Tata}, i.e. the space of $n$ soliton sections
of some $su(n)$ bundle.

Now we turn to show that the $W_1^{\alpha_1}, W_2^{\alpha_2}$ with
$(\alpha_1, \alpha_2) \in Z_n \times Z_n$ acting on this ${\cal H}_n$
generates the N.C. algebra ${\cal A}_n$ on ${\cal T}_n$.

\subsection{Noncommunitative algebra ${\cal A}_n$ on fuzzy torus ${\cal
T}$ as the Heisenberg Weyl group $Z_n \times Z_n$}

\indent

From section 2 we know that the effects of the noncommutative Wilson loop
\cite{AMNS, BA} $W_1 = U_1^{\frac{1}{n}}$ and $W_2 = (U_2^{\frac{1}{n}})$
on the $i$-th soliton is to translate its position, from $z_i$ to $(z_i +
\frac{1}{n} \tau - \delta_{in}\tau)$ and $(z_i + \frac{1}{n} -
\delta_{in})$ respectively, or equivalently shift all $z_i$ by
$\frac{1}{n}$ or $\frac{\tau}{n}$ mod the torus ${\cal T}$, furthermore
equivalent to shift the coordinate origin $u$ by $\frac{1}{n}$ of
$\frac{\tau}{n}$ in {\it opposite direction}. Substituting in
(\ref{basis}), we find
\begin{eqnarray}
\label{funcform-w1}
\hat{\bf V}_1 v_a(z_1, \cdots, z_n) & = & (\prod_{i = 1}^{n - 1}
T^{(i)}_{\frac{\tau}{n}}) T^{(n)}_{\frac{\tau}{n} - \tau}
v_a(z_1, \cdots, z_n)\\
\label{matrixform-w1}
& = & v_{a - 1}(z_1, \cdots, z_n) = W_1 v_a(z_1, \cdots, z_n),\\
\label{funcform-w2}
\hat{\bf V}_2 v_a(z_1, \cdots, z_n) & = & v_a(z_1 + \frac{1}{n},
\cdots, z_n + \frac{1}{n} - 1)\\
\label{matrixform-w2}
& = & (-1)^{n+1}e^{2\pi i \frac{a}{n}} v_a(z_1, \cdots, z_n) = W_2
v_a(z_1, \cdots, z_n),
\end{eqnarray}
where
\begin{equation}
T^{(i)}_{a \tau} f(z) \equiv e^{\pi i a^2 \tau + 2 \pi i a z_i}f(z_1,
\cdots, z_i + a \tau, \cdots, z_n).
\end{equation}

Using basis $v_a$ (\ref{basis}), we have transform the action of the N.C.
Wilson loop of ${\cal T}_n$ from the shift in functional space form
(\ref{funcform-w1}) and (\ref{funcform-w2}) into the matrix operator form
(\ref{matrixform-w1}) and (\ref{matrixform-w2}), i.e.
\begin{equation}
\label{comm-rule}
(\hat{\bf V}_1)_{a b} = \delta_{a + 1, b}; \hspace{.5cm}
(\hat{\bf V}_2)_{a b} = \delta_{a b} \omega^{a}, \hspace{.5cm}
\omega = e^{\frac{2\pi i}{n}} = e^{i \theta_n}
\end{equation}
As expected this $\hat{\bf V}_1$ and $\hat{\bf V}_2$ satisfy the relations
\begin{equation}
\hat{\bf V}_1^\dagger \hat{\bf V}_1 = \hat{\bf V}_2^\dagger \hat{\bf V}_2
= 1, \quad \hat{\bf V}_1^n = \hat{\bf V}_2^n = 1, \quad \hat{\bf V}_1
\hat{\bf V}_2 = \hat{\bf V}_2 \hat{\bf V}_1 e^{2\pi i\theta_n},
\end{equation}
here $\theta_n = \frac{1}{n}$ is the noncommutative parameter for the $n$
2-branes \cite{SW}.

Remark: There are some subtle and crucial points in our paper which is
lialbe to cause confusion. It seem worth to be stressed and clarified
here. Originally for ${\cal H}_{\cal T}$ on total ${\cal T}$ we have the
commuting $U_1$ and $U_2$ with $\theta = 2 \pi Z$. After orbifolding, it
is clear \cite{CDS, SW} that ${\cal H}_{\cal T} = V_n \otimes {\cal L}$ is
acted by $U_i = W_i \otimes \hat{\bf V}_i^{-1}$, where the matrices $W_i$
satisfy $W_1 W_2 = W_2 W_1 \omega$, while the difference operators
\begin{equation}
\hat{\bf V}_1^{-1} \hat{\bf V}_2^{-1} = \hat{\bf V}_2^{-1} \hat{\bf
V}_1^{-1} e^{\theta - \frac{2 \pi i}{n}} \equiv \hat{\bf V}_2^{-1}
\hat{\bf V}_1^{-1} e^{- 2 \pi i \theta_n}.
\end{equation}
Generally $W_i^\alpha$ is represented by the $Z_n \times Z_n$ matrices
$I_\alpha$ on trivial $su(n)$ bundle $V_n$, while $\hat{\bf
V}_\alpha^{-1}$ is the translation of the origin of ${\cal T}$: $u = 0
\longrightarrow u = - \frac{\alpha_1 + \alpha_2 \tau}{n}$,
correspondingly, to shift the center of mass of $n$ solitons $z_{\rm c} =
\frac{1}{n} \sum_{i = 1}^n z_i$ by $\frac{\alpha_1 + \alpha_2 \tau}{n}$.
If we restrict to consider the ${\cal H}_n = V_n$ on subtorus ${\cal
T}_n$, then we may either realize the $W_i^\alpha$ in operator form i.e.
in terms of $Z_n \times Z_n$ matrix (\ref{matrixform-w1})
(\ref{matrixform-w2}) or realize it in functional formalism i.e. in terms
of difference operators (\ref{funcform-w1}) (\ref{funcform-w2}). These two
forms is related by the Weyl Moyal transformation on torus. But it happens
that the $\hat{\bf V}_\alpha$ ($z_{\rm c} \longrightarrow z_{\rm c} +
\frac{\alpha_1 + \alpha_2 \tau}{n}$) yield a matrix transform on our basis
function, so it is equivalent to the  operator form of $W_\alpha$. Thus
for the total covering torus ${\cal T}$ the $W_\alpha$ and $\hat{\bf
V}_\alpha^{-1}$ cancels, $U_\alpha = W_\alpha \otimes \hat{\bf
V}_\alpha^{-1}$ are commutative, as the (6.54) in \cite{SW}.

The functional and the operator forms are related by Weyl Moyal
transformations on torus (see Appendix B of \cite{GHS}), functions $v_a$
behaves as the coherent "symbols", Fock-Bargmann representation functions
\cite{Perelomor} on torus.

On the space ${\cal H}_n = \{ v_a | a = 1, 2, \cdots, n \}$, the {\it
noncommutative algebra ${\cal A}_n$ generated by $W_1$ and $W_2$} has been
truncated and becomes a $n \times n$ dimensional unital $C^*$ algebra with
$n^2$ basis $W_1^{\alpha_1} W_2^{\alpha_2} = W^{\alpha}$ and the define
relations of ${\cal A}_n$
\begin{equation}
\label{w-comm}
W^{\alpha \beta}W^{\alpha \beta \dagger} = 1, \quad W^{\alpha
\beta} W^{\alpha^\prime \beta^\prime} = W^{\alpha^\prime \beta^\prime}
W^{\alpha \beta} \omega^{\alpha \beta^\prime - \beta \alpha^\prime} =
W^{\alpha + \alpha^\prime, \beta + \beta^\prime}\omega^{-\beta
\alpha^\prime}, \quad (W^{\alpha \beta})^n = 1.
\end{equation}
Here and hereafter to represent in matrix (operator) form the basis
$W^{\alpha_1, \alpha_2}$ of ${\cal A}_n$ we introduce the usual $Z_n
\times Z_n$ matrix $(I_\alpha)_{a b} = (I_{\alpha_1, \alpha_2})_{a b} =
\delta_{a + \alpha_1, \alpha_2}\omega^{b \alpha_2}$ \cite{Tata}. This
algebra in difference operator form is the Heisenberg group $G_{\cal
H}(n)$ $Z_n \times Z_n$ in ordinary $\theta$ function theory \cite{Tata},
including both the shift $\frac{\alpha \tau}{n} + \frac{\beta}{n}$ of
arguments and the change of phases.

\subsection{$su(n)$ algebra $g$ on N.C. torus}

\indent

It is shown \cite{CFH} that the level $l$ representation of the Lie
algebra $sl_n({\cal T})$ on the elliptic curve ${\cal T}$ can be written
as following:
\begin{equation}
\label{E_alpha}
E_{\alpha} = (-1)^{\alpha_1}\sigma_{\alpha}(0)\sum_j\prod_{k \neq j}
\frac{\sigma_{\alpha}(z_{jk})}{\sigma_0(z_{jk})}\left [ \frac{l}{n}
\sum_{i \neq j}\frac{\sigma_{\alpha}^\prime(z_{ji})}
{\sigma_{\alpha}(z_{ji})} - \partial_j \right ],
\end{equation}
and
\begin{equation}
\label{E_0}
E_0 = -\sum_j \partial_j,
\end{equation}
here $\alpha \equiv (\alpha_1, \alpha_2) \in Z_n \times Z_n$ and $\alpha
\neq (0, 0) \equiv (n, n)$, $z_{jk} = z_j - z_k$, $\partial_j =
\frac{\partial}{\partial z_j}$. $E_0$ commutes with $E_{\alpha}$,
$sl_n({\cal T})$ includes only the $E_{\alpha}$ with $\alpha \neq 0$.
After a complicate calculation, we obtain the commutation relation:
\begin{equation}
\label{E-comm}
[E_{\alpha}, E_{\gamma}] = (\omega^{-\alpha_2 \gamma_1} -
\omega^{-\alpha_1 \gamma_2}) E_{\alpha + \gamma},
\end{equation}
or in usual $sl(n)$ with $i, j$ label the Chan Paton indices basis, let
\begin{equation}
\label{E_ij}
E_{ij} \equiv \sum_{\alpha \neq 0} (I^\alpha)_{ij} E_{\alpha},
\end{equation}
then
\begin{equation}
[E_{jk}, E_{lm}] = E_{jm}\delta_{kl} - E_{lk}\delta_{jm}.
\end{equation}
Remark: The representation (\ref{basis}) and the commuation rules
(\ref{E-comm}) can also be obtained \cite{CFH} from a quasiclassical limit
from the representation of the $Z_n \times Z_n$ Sklyanin albebra
\cite{Sklynin}. Sklyanin and Takebe \cite{ST} give the elliptic $sl(2)$ by
using double periodic Weierstrass functions. The high spin $l$
representations is given also. In this paper, it is restricted to the $l =
1$ representation of $sl_n({\cal T})$ by holomorphic sections on
${\cal T}^{\otimes n} / S_n$.

\subsubsection{ Automorphism of $E_{\beta} \in su_2({\cal T})$ by
noncommutative gauge transformation $W^{\alpha} \in {\cal A}$}

\indent

Since the Wilson loops $W_1$ and $W_2$ acting on the noncommutative
covering torus ${\cal T}$ is to shift $z_i$ to $(z_i + \frac{\tau}{n} -
\delta_{in}\tau)$ and $(z_i + \frac{1}{n} - \delta_{in})$ respectively, so
$E_{\alpha}$ in (\ref{basis}) will be changed into
\begin{eqnarray}
\label{wE_alpha}
W_1\cdot (E_\alpha(z_i)) & = & W_1 E_\alpha(z_i) W_1^{-1}  =
E_\alpha(z_i + \frac{\tau}{n} - \delta_{in}\tau) =
\omega^{-\alpha_2}E_\alpha(z_i),\nonumber\\
W_2\cdot (E_\alpha(z_i)) & = & W_2 E_\alpha(z_i) W_2^{-1} =
E_\alpha(z_i + \frac{1}{n} - \delta_{in}) =
\omega^{\alpha_1}E_\alpha(z_i).
\end{eqnarray}
or more generally
\begin{equation}
\label{automorphism}
W^{\beta_1, \beta_2}\cdot E_\alpha = W^{\beta_1, \beta_2} E_\alpha
(W^{\beta_1, \beta_2})^{-1} = \omega^{\alpha_1\beta_2 - \alpha_2\beta_1}
E_\alpha.
\end{equation}
This could be compared with the matrix model on noncommutative torus
\cite{CDS}, where
$$
U_i X_j U_i^{-1} = X_j + \delta_{ij} 2 \pi R_j
$$
or more exactly with the covariant derivatives
$$
U_i \bigtriangledown_j U_i^{-1} = \delta_{ij}\bigtriangledown_j
$$
i. e.
$$
E_\alpha \Longrightarrow \exp(\frac{i \pi}{n}
\bigtriangledown_1)^{\alpha_1} \exp(\frac{i \pi}{n}
\bigtriangledown_2)^{\alpha_2}.
$$

\subsubsection{Isomorphism of $su_n({\cal T})$ and ${\cal A}_n$ on
${\cal H}_n$}

\indent

let $E_\alpha \in g$  to act on $v_a$, we find that
\begin{equation}
\label{E-effect}
E_\alpha v_a = \sum_b(I_\alpha)_{b a}v_b.
\end{equation}
As in section 3.2 we already learn that $W^\alpha v_a = \sum
(I_\alpha)_{ba}v_b$, so on ${\bf\cal H}_n$ we establish the isomorphism
\begin{eqnarray}
\label{isomorphism}
su_n({\cal T}) & \Longrightarrow & {\bf\cal A}\nonumber\\
E_\alpha & \longrightarrow & W^\alpha
\end{eqnarray}

Obviously this is correspondent with the noncommutative plane case, where
one have the homomorphism of the operators $\partial$ and $[ \quad,*
\quad]$
\begin{equation}
\label{f-corp}
i\epsilon_{ji}\partial_{x_i} F(x) \Longrightarrow [x, * f(x)]
\end{equation}
and the isomorphism upon acting on Fock space ${\cal H}$
\begin{equation}
\label{op-corp}
\partial_z \longrightarrow a, \hspace{1cm} \partial_{\bar{z}}
\longrightarrow a^\dagger
\end{equation}
Compare the eq. (\ref{f-corp}) with the eqs. (\ref{wE_alpha})
(\ref{automorphism}) and compare the eq.(\ref{op-corp}) with the eq.
(\ref{E_alpha}) and (\ref{E-effect}), it is easy to see that the
infinitesimal translations $\partial_i$ on plane ${\bf R}^2$ corresponds
to $E_\alpha$ of the $su_n({\cal T})$ and the algebra ${\cal A}_{{\bf
R}^2}$ of $a, a^*$ corresponds to the algebra ${\cal A}_n$ of $W_i$.
Meanwhile the local adj operation for ${\cal A}_{{\bf R}^2}$
(\ref{f-corp}) changes to the global Adj operation (gauge transformation
\cite{AMNS, BA}) for ${\cal A}_n$ (\ref{wE_alpha}).

\subsubsection{$\hat{V}_i$ and $E_\alpha$ as generators of the Weyl
reflection group}

\indent

Kac \cite{Kac} has shown that the $\theta$ function with characteristic
$\alpha = (\alpha_1, \alpha_2) \in Z_n \otimes Z_n$ transforms under
affine Wely reflections as the Heisenberg group generated by $\hat{V}_i$
(\ref{funcform-w1}), (\ref{funcform-w2}). As we have shown in
(\ref{matrixform-w1}), (\ref{matrixform-w2}) and (\ref{E-effect}), both
$\hat{V}_\alpha$ and $E_\alpha$ act on $v_a$ as $I_\alpha$, operators
$\hat{V}_1$ and $E_1$ as the coxeter element $I_1$, operators $\hat{V}_2$
and $E_2$ as the cyclic element $I_2$; operator $E_{i j} (i \neq j)$
permutes $i, j$ i.e. gives the reflection $\hat{r}_{i j}$ reversing the
root $e_i - e_j$.

Here, let us stress the relations of the geometrical N.C. algebras ${\cal
H}_n$, ${\cal A}_n$ and $su(n)$ with the physical properties of N.C.
solitons. Obviously, since in (\ref{basis}) each term $F_\alpha$ are
symmetrical with respect to permutation $S_n$ of the centers $z_i$ of
solitons, so the vectors $v_\alpha$ is defined on the module ${\cal
T}_n^{\otimes n}/S_n$. The operator $\hat{\bf V}_1$ shifts each soliton
$z_i$ from the $i$-th covering of ${\cal T}$ over ${\cal T}_n$ i.e. from
the $i$-th brane to the next one, or equivalently changes each brane to
the previous one. The operator $\hat{\bf V}_2$ shifts the $U(1)^{\otimes
n}$ phase of $i$-th brane by $\omega^i$. The Chan Paton $su(n)$ of $n$
branes is generated by $E_{i j}$. Acted on ${\cal H}_n$, both $\hat{\bf
V}_i$ and $E_{i j}$ generate the Weyl reflection. The offdiagonal elements
give permutations of the branes, the diagonal elements generates the phase
shifts of soliton wave functions.

\section{Affine algebra on N.C. torus, integrable elliptic Gaudin \\
and Calogero Moser models \cite{CFH}}

\indent

Now we will show that the cotangent bundle for N.C. elliptic $su_n({\cal
T})$ (\ref{E_alpha}) with twisted loop $W_i$ realizes the elliptic Gaudin
model on N.C. torus, which is related to critical level twisted $su(n)$
WZW i.e. $A_{n - 1}$ affine algebra on N.C. torus. Then use the elliptic
Vandermounde determinant to gauge transform it into elliptic C.M. model
which is the Hamiltonian reduction of the cotangent bundle for the algebra
of the semidirect product of the Cartan torus and the Weyl group
\cite{HM}.

\subsection{N.C. Elliptic Gaudin Model}

\indent

The elliptic Gaudin model on commutative space \cite{ST1} is defined by
the transfer matrix (quantum lax operator):
\begin{equation}
\label{L(u)}
L_{i j}(u) = \sum_{\alpha \neq (0, 0)} w_\alpha (u) E_\alpha(I_\alpha)_{i
j},
\end{equation}
where
\begin{equation}
\label{w_alpha(u)}
w_\alpha(u) = \frac{\theta^\prime(0) \sigma_\alpha(u)}
{\sigma_\alpha(0) \sigma_0 (u)},
\end{equation}
$E_\alpha$ and $I_\alpha$ are the generators of $su(n)$ (or ${\cal
A}_{n-1}$ Weyl) and $G_{\bf\cal H}(n)$ in the quantum and auxiliary space
respectively. Using the general defining relations of $su(n)$
(\ref{E-comm}) and $G_{\cal H}(n)$ (\ref{w-comm}), we find that the
commutators of $L$ (quantum version of fundamental Poisson bracket):
\begin{equation}
\label{L-comm}
[L^{(1)}(u_1), L^{(2)}(u_2)] = [r^{(1, 2)}(u_1, u_2), L^{(1)}(u_1) \oplus
L^{(2)}(u_2)],
\end{equation}
where the classical Yang-Baxter matrix
\begin{equation}
\label{r(u)}
r(u)_{i, j}^{k, l} = \sum_{\alpha \neq (0, 0)}w_\alpha(u) (I_\alpha)_i^k
\otimes (I_\alpha^{-1})_j^l
\end{equation}
is antisymmetrical
\begin{equation}
r_{i, j}^{k, l}(u) = r_{j, i}^{l, k}(-u)
\end{equation}
and satisfies the classical Yang-Baxter equation:
\begin{equation}
[r_{1 2}(u_{1 2}), r_{1 3}(u_{1 3})] + [r_{1 2}(u_{1 2}), r_{2 3}(u_{2 3}]
+ [r_{1 3}(u_{1 3}), r_{2 3}(u_{2 3}] = 0,
\end{equation}
where $u_{i j} = u_i - u_j$. Thus this system is integrable. Indeed, we
can find from the quantum determinant of $L$ the complete set of
Hamiltonian. Some examples may be found in \cite{ST1, ER, KT1}. It was
obtained also as the nonrelativistic limit of the Ruijsenaars-Macdonald
operators which has been described in the part two of our unsubmitted
paper \cite{HPSY}. As a lattice model, the common eigenfunctions and
eigenvalues of Gaudin model is solved in terms of Bethe ansatz \cite{FFR},
which has been expressed by the conformal blocks of the twisted WZW models
on the torus \cite{FW, KT}.

To relate these well known results about this usual Gaudin model with this
newly introduced Gaudin model on the fuzzy torus ${\cal T}_n$ (section 3),
let us substitute the differencial representation (\ref{E_alpha}) of
$su_n({\cal T})$ $E_\alpha$ into the Gaudin $L$ (\ref{L(u)}), then the $L$
turns to be in a factorized form
\cite{CFH}:
\begin{equation}
\label{factorized_L}
E_0 + \sum_{\alpha \neq (0, 0)}w_\alpha(u) E_\alpha (I_\alpha)_j^i =
L(u)_j^i = \sum_k \phi(u, z)_k^i \phi^{-1}(u, z)_j^k \partial_u - l \sum_k
\partial_u \phi(u, z)_k^i \phi^{-1}(u, z)_j^k ,
\end{equation}
where the factors $\phi^i_k$ are the vertex face intertwiner
\begin{equation}
\label{intertwinner-factors}
\phi(u, z)_j^i = \theta_{\frac{1}{2} - \frac{i}{n}, \frac{1}{2}}(u + n z_j
- \sum_{k = 1}^n z_k + \frac{n-1}{2}, n \tau).
\end{equation}
We can also show that (\ref{factorized_L}) is a representation of $L$ by
subsituting it directly into (\ref{L-comm}). Now one may find that the
expression of each elements of $r$-matrix becomes
\begin{equation}
\label{intertwinner-factor}
r_{i, j}^{l, k}(u) = \delta_{i + j}^{l + k} \left \{ (1 - \delta_i^l)
\frac{\theta^0(0)^\prime \theta^{i - j}(u)}{\theta^{l - j}(u) \theta^{i -
l}(u)} + \delta_i^l \left ( \frac{\theta^{i - j}(u)^\prime}{\theta^{i -
j}(u)} - \frac{\theta(u)^\prime}{\theta(u)} \right ) \right \},
\end{equation}
where $\theta^i(u) \equiv \theta_{\frac{1}{2} - \frac{i}{n},
\frac{1}{2}}(u, n \tau)$. It is easy to find that it is $Z_n \otimes Z_n$
symmetrical, $r(u) = (I_\alpha \times I_\alpha) r(u) (I_\alpha^{-1}
\otimes I_\alpha^{-1})$ and turns to be equal to the sum in (\ref{r(u)}),
i.e. actually the factorized differential operators $L_j^i$
(\ref{factorized_L}) realizes a representation of the defining relation
(\ref{L-comm}).

The intertwinner "factor"s (\ref{intertwinner-factors}) , now intertwines
the Chan Paton $su(n)$ index $i$ for the brane (vertex model index) with
the dynamical indices $j$ (face model indices) of dynamical soliton
position $z_j$ on the world sheet.

Feigin et. al \cite{FFR} has established the relation between the critical
level $su(n)$ WZW models and the rational Gaudin model, that is the
expressions of $L(u)$ and $R(u)$ being similar as (\ref{L(u)}) and
(\ref{r(u)}), but instead of the elliptic $w_\alpha(u)$ in
(\ref{w_alpha(u)}) now with a rational $w_\alpha^R(u) = \frac{1}{u}$, $u
\in {\bf C}$. Kuroki and Takebe \cite{KT} find the same relation of
$RLL$ for the elliptic case, and point out that the WZW is twisted, i.e.
is defined on a twisted bundle
$g^{\rm tw}$:
$$
g^{\rm tw} := ({\bf C} \times g)/\sim
$$
where the equivalence relation $\sim$ are $(u, g) \sim (u + 1, I_1 g
I_1^{-1}) \sim (u + \tau, I_2 g I_2^{-1})$. Correspondingly the
$w_\alpha(u)$ (\ref{w_alpha(u)}) is a meromorphic section of the twisted
bundle with single pole at $u = 0$ and residue $1$. While wrapping around
different one cycle ($+1$ or $+\tau$) on the base torus, the global gauge
transformation $U_1$ or $U_2$ for sections in the fibre are
noncommutative, but the sections are holomorphic functions of $u$ on base
torus, which has only a complex conformal structures, it has neither
metric nor symplectic structure, thus remains to be commutative. For our
N.C. torus we have both $z = y_1 + i y_2$ and $\bar{z} = y_1 - i y_2$
which are N.C., i.e. $y_i$ are orthomormal with respect to the metric and
N.C. symplectic at the same time, it causes both the local N.C. $z,
\bar{z}$ and the global N.C. $W_1, W_2$.

For the usual Gaudin \cite{ST1, ER} on a marked commutative torus, the
$z_i$ is the marked points, i.e. the poles of $L(u) = \sum_i \sum_\alpha
w_\alpha(u - z_i) E_\alpha I_\alpha$. The $p_i$ is the conjugate momentum:
classically $\{p_i, q_j\} = \delta_{i j}$, quantumly $[p_i, q_j] =
\delta_{i j}$. So, the $L$-matrix is endowed with dynamics, by given the
Poisson (equivalently Konstant-Kirrilov) brackets at first, then
by quantized the $p_i \sim -i \frac{\partial}{\partial q_i}$ to find the
matrix differential form of $L$ \cite{CFH}. While now for the Gaudin on
noncommutative torus, the $z_i$ are the center (position) of the $i$-th
soliton, $\partial_i$ as its infinitesimal translation is equivalently to
the $[\bar{z}_i *, ]$. Here we should stress that the key point to
translate the usual integrable models on commutative space to that on N.C.
space is to notice that since in N.C. space $[\bar{z}_i, z_j] = \delta_{i
j}$, so N.C. plane automatically become the symplectic manifold
corresponding to the quantum phase space. $p_i \rightarrow \bar{z}_i$,
$q_i \rightarrow z_i$ i.e in the holomorphic Fock Bargmann formalism
$[\bar{z}, f]$ automatically becomes $\partial_z f$ just like $p
\rightarrow \frac{\partial}{\partial q}$. Apparently, the complex
structure with given metric determines naturally a symplectic structure,
moreover it is automatically "quantized", as in the geometrical
quantization.

Let us show, that the dynamical equation, induced by N.C. this way,
really gives the interaction between N.C. solitons assumed in \cite{GHS},
by studying the potential and the spectral curves of C.M. model.

\subsection{Elliptic Calogero Moser model and its equivalence with
elliptic Gaudin model}

\indent

The elliptic C.M. model is defined by the Himiltonian:
\begin{equation}
\label{CM-Hamiltonian}
H = \sum_{i = 1}^n \partial_i^2 + \sum_{i \neq j} g \wp(z_{i j}),
\end{equation}
where $\wp(z) = \partial^2 \sigma(z)$. This quadratic and other higher
Hamiltonian are generated by the Krichver Lax matrix
\begin{equation}
L_{\rm Kr}(u)^i_j = \partial_i \delta_j^i + (1 - \delta_j^i) \sqrt{g}
\frac{\sigma(u + z_{j i})}{\sigma(u) \sigma(z_{j i})}.
\end{equation}
where $z_{j i} = z_j - z_i$. By the Poisson transformation (classically
$\partial_i \sim p_i$, $z_i \sim q_i$)
\begin{equation}
p_i \longrightarrow p_i - \frac{\partial}{\partial q_i} \ln
\Pi^{\frac{l}{n}}(q),
\end{equation}
here
\begin{equation}
\Pi(q) \equiv \prod_{i < j}\sigma(q_{i j}), \hspace{1cm} \sqrt{g} = -
\frac{l}{n} \sigma^\prime(0),
\end{equation}
and $L_{\rm Kr}(u)_j^i$ becomes
\begin{equation}
\label{L_CM}
L_{\rm CM}(u)_j^i = (p_i - \frac{l}{n}\frac{\partial}{\partial q_i}\ln
\Pi(q))\delta_j^i - \frac{l}{n}\sigma^\prime(0)(1 - \delta_j^i)
\frac{\sigma(u + q_{j i})}{\sigma(u) \sigma(q_{j i})}.
\end{equation}
This may be further gauge transformed into the factorized $L$
(\ref{factorized_L}) of Gaudin model by gauge transformation matrix
\begin{equation}
\label{gaugem}
G(u; q)_j^i \equiv \frac{\phi(u; q)_i^i}{\prod_{l \neq j}\sigma(q_{j
l})}, \quad \mbox{no summation of $i$ in $\phi_i^i$}.
\end{equation}

It is well known that C.M. model gives the {\it dynamics} of a long
distance interaction between $n$-bodies located at $z_i$, $(i = 1, 2,
\cdots, n)$. Now we will show that it is the interaction between
$n$-solitons on the fuzzy torus while $z_i$ becomes the positions of the
centers of each soliton. According to paper \cite{GHS}, for N.C.
multisolitons the potential term is argumented to be the Laplacian of a
K\"ahler potential ${\cal K}$, which is the logrithm of a Vandermonde
determinant. Actually we have
\begin{eqnarray}
\sum_{i \neq j}\wp(z;j) & = & \sum_i \partial_i^2 \log \prod_{j \neq k}
\sigma(z_j - z_k) \equiv \sum_i \partial_i^2 {\cal K}(u, z),\\
e^{{\cal K}(u, z)} & = & \prod_{j \neq k} \sigma(z_j - z_k)\sigma(n u +
\frac{n(n - 1)}{2})\nonumber\\
& = & \det(\phi_k^j) \equiv \sigma(n u + \frac{n(n - 1)}{2})\prod_{i \neq
j}\sigma(z_{i j}).
\end{eqnarray}
The variable $u$ of the marked torus is the so called spectral parameter
or evaluation parameter of Lax matrix $L_{\rm Kr}(u)_j^i$.

As \cite{BCS}, replacing the "spin" index $i, j$ elements $E_{i j}$
(\ref{E_ij}) in the Lax matrix $L_{\rm Kr}(u)_j^i$ of the C.M. model by
soliton exchange $z_i \leftrightarrow z_j$ permutation $s_{i j}$ in ${\cal
A}_{n-1}$ Weyl reflection, $L_{\rm Kr}(u)_j^i$ becomes like the Dunkle
operators \cite{BFV}. This "spin" index exchange equivalence with particle
(soliton) exchange is obtained if restricted to be acted on the space of
wave functions which is totally (anti)symmetric in both spins and
positions. This is the {\it ${\cal A}_{n-1}$ Weyl symmetry} for the
$su_n({\cal T})$ Chan Paton index of the branes and for the positions
$z_i, z_j$ of solitons.

Here the determinant of the vertex-face interaction is the determinant of
our gauge transformation matrix from Gaudin to C.M. We should mention that
its $u$ independent factor $\prod(z)$ is the Weyl antiinvariant ground
state wave function of C. M. It is the "phase functions" part of the
conform block of the twisted WZW model on elliptic curve related to the
Gaudin model \cite{FFR, FW, KT1}. Dividing by this phase the KZB equation
of the elliptic Gaudin Model reduces \cite{FW, KT1} to the heat equation
associated to the elliptic C. M. equation.

\section{C. M. and Hitchin systems. Moment map and BPS equation,
Spectral curves and brane configuration}

\indent

\subsection{Hitchin system for the cotangent bundle of $W \otimes
{\cal T}_n$ \cite{HM}}

\indent

Instead as in section 4.2, where we obtained $L_{\rm CM}$ from $L_{\rm
Gaudin}$, now the C.M. $L_{\rm CM}$ operators (\ref{L_CM}) as the Weyl
reflection of solitons \cite{BCS} may be derived directly as a Hitchin
systems for a group $G$ on root space. $G$ is an extension of the
semidirect product of the Weyl group and the Cartan maximum torus of the
roots. The cotangent bundle of $G$ consists of $su(n)$ connection $A$ and
Hermitean $n \times n$ matrix forms scalar field $\varphi$.

The {\it moment map} condition: for $A$, the transition functions on root
part (offdiagonal) is zero; for root part of $\varphi$ ($\varphi_{\rm
offdiagonal}$) is: ${\rm res}_{u = 0}(\varphi_{\rm offdiagonal}) = $ a
Weyl invariant constant $\sqrt{g}$. The solution turns to be
\begin{eqnarray}
A & = & {\rm diag}(q_1, q_2, \cdots, q_n)\\
\label{varphi_ij}
\varphi(u)_j^i & = & p_i \delta_j^i + (1 - \delta_j^i)\sqrt{g}
\frac{\sigma(u + q_{j i})}{\sigma(u) \sigma(q_{j i})},
\end{eqnarray}
here $\{p_i, q_j\} = \delta_{i j}$.

Correspondingly for our case of N.C. soliton, this classically
Poisson conjugate $p_i$ and $q_i$ in phase space should be automatically
quantized and be replaced by quantum $\partial_i$ ($\sim \bar{z}_i$) and
$z_i$ with $[\partial_i, z_j] = \delta_{i j}$ ($\sim [\bar{z}_i, z_j] =
\delta_{i j}$ on N.C. ${\cal T}$)and (\ref{varphi_ij}) becomes the matrix
differential operator
\begin{equation}
\varphi(u)_j^i = L_{\rm Kr}(u)_j^i.
\end{equation}

\subsection{Brane picture}

Kapustin and Sethi \cite{KSe} consider the case of a single D4 wrapped on
${\cal T}^2$ probing by $n$ D-0 branes with FI (Fayet-Iliopoulos)
deformation. The Higgs branch is given by the moduli space of solutions of
the {\it D-flat condition}, but with impurity source (moment map equation
Douglas Moore \cite{DM})
\begin{equation}
\label{mapeq}
\partial \varphi + [A_\mu, \varphi] = \zeta ( 1 - n | v \rangle \langle v
|) \delta^2 (u), \quad \langle v | = \frac{1}{\sqrt{n}}(1, 1, \cdots, 1),
\end{equation}
here the $| v \rangle \langle v |$ term comes from the fundamental
hypermultiplet of the single 0-4 string, $\zeta \sim$ the FI deformation
parameter. Remark: The paper \cite{KSe} consider the case of more 0-4
strings (impurities), such that the r.h.s. of (\ref{mapeq}) includes the
sum of $\sigma^2(u_i)$, which will corresponds to the marked points $u_i$
on torus \cite{ER}, to the lattice points $u_i$ for Gaudin models
\cite{FFR}, or to the vertices for conformal block and KZB equations
\cite{FW, KT1}.

Since the $n$ D0 branes dissolved into ${\cal T}$ as the B-flux, the
torus becomes N.C. \cite{BBST}. In fact, recently, Susskind et al.
\cite{Susskind, Polychronakos, GKK} proposing the noncommutative
Chern-Simons theory of the QHE on 2-brane, use this equation as a
constrain equation (moment map equation). Here $ | v \rangle \langle v |$
comes from the boundary field \cite{Polychronakos} or equivalently as the
Wilson loop \cite{MP} related to the quasiparticle excited states.

Compare with \cite{Susskind, Polychronakos} we see that the commutator
equation (the first one in (\ref{w-comm})) of $U_i$ generating the ${\cal
A}_n$ algebra in section 3 generalizes their {\it vortex free} ({\it
ground state}) constrain equation, while the elliptic C. M. moment map
equation (\ref{mapeq}) generalizes their {\it constrain equation with
vortex} ({\it quasiparticle}) from their case of $N \subset R^2$ to
our case of N. C. torus.

Before study the QHE in next section, let us sketch the {\it algebraic
geometrical} picture of the {\it branes}.

Krichver \cite{Krichver} defined the {\it spectral curve} $\Gamma$ of
elliptic C.M. model by
\begin{equation}
\label{Gamma}
\Gamma(k, u) \equiv \Gamma_{p, q}(k, u) \equiv {\rm det}|k - L_{\rm
Kr}(u)^i_j| = \prod (k - K_i(u, p, q) = 0,
\end{equation}
where the spectral parameter $u \in \Sigma_0 = \frac{C}{Z + Z \tau}$. The
characteristic function $\Gamma(k, u)$ is an order $n$ polynomial of $k$,
the coefficients or the eigenvalue $K_i(u, p, q)$ for any given $u$
depends on the modules $p_i, q_i$. The explicit solution of $K_i$ has been
given by D'Hoker and Phong \cite{DP1, DP2}. The curve $\Gamma$ covers the
bare $\Sigma_0$ $n$ times. Upon the marked point $u=0$, $K_i(u) \sim
\frac{1}{u}, (i = 1, \cdots, n-1)$, $K_n(u) \sim \frac{1-n}{u}$, for
generic $u$, $\Gamma(k, u)$ has $n$ zeros $K_i$ seperately on the $i$-th
sheet. The number of quadratic branch points as the zeros of
$\frac{\partial \Gamma}{\partial k}$ is $2(n - 1)$ i. e. the genus of
$\Gamma$ is $n$.

In Witten's construction of {\it DHWW brane} \cite{D, HW, Witten} via M
theory, the components on the worldvolume of the brane wrapping on a
compactified target spans a two dimensional surface $\Sigma$ embeding in a
4 dimensional manifold $Q(x_4, x_5, x_6, x_{10})$.

The $N = 2$ supersymmetry gives $Q(k, u)$ the complex structure $k = x_4 +
i x_5$, $u = x_6 + i x_{10}$, then $\Sigma(u)$ is a Riemann surface. While
the 4-manifold $Q$ is the moduli space of the ALF metric generated by $n$
parrallel $6$-branes transverse to $Q$. The ALF metric (e.g. Taub-Nut
metric), being a hyperquotient, endows the manifold $Q$ a complex
symplectic structure \cite{GR}.

Donagi and Witten \cite{DW, Witten} find that for the integrable complex
Hamiltonian systems the $\Sigma$ will move in $Q$ such that there exist a
line bundle $L_{\Sigma}(u)$ on $\Sigma(u)$, which can be deformed into
$L_{\Sigma}(u^\prime)$ on $\Sigma(u^\prime) \subset Q$, $(u \equiv
u(t_0), u^\prime \equiv u(t_0 + \epsilon))$, by the Hamiltonian flow from
time $t_0$ to $t_0 + \epsilon$. If we find a brane such that its $\Sigma$
in $Q$ has the bundle section $L_{\Sigma}$ equals the Lax matrix $L_{\rm
Kr}$ of C.M., then the SW curve $\Sigma(k, u)$ of this brane
configuration can be identified with the spectral curve $\Gamma(k, u)$ of
integrable C.M. systems in the following way. Since the D flat equation
(\ref{mapeq}) \cite{KSe} (or BPS like equation \cite{MMM}) for brane or
SUSY \cite{KSe, DM, MMM} is the same as the C.M. moment map equation,
which reduces the $sl_n({\cal T})$ cotangent bundle into the ellliptic
C.M., so the solutions of the D flat equation, the embedding target field
$\varphi(u)^i_j$ matrix of brane, corresponds to the offdiagonal element
of the C.M. Lax operator $L_{\rm Kr}(u)^i_j$. The $L_{\rm Kr}(u)$ is the
section of the bundle on the spectral curve $\Gamma(k, u)$ after pulled
back $u$ on $\Sigma_0$ by (\ref{Gamma}) to $(k, u)$ on $\Gamma(k, u)$.
Thus the spectral curve $\Gamma(k, u)$ is identified with the two
dimensional surface $\Sigma(k, u)$ ("target direction" $x_4 + i x_5 = k$,
"world sheet" $x_6 + i x_{10} = u$) wrapping the target brane. Thus, the
map from the SW $\Sigma$ to the spectral $\Gamma$ is a symplectic map
\cite{K, KP} from the SW modules to the C.M. modules $p_i, q_i$.

Here it is crucial that the world sheet coordinates $x_6, x_{10}$ of base
torus become noncommutative by the $B$ flux, consequently $p_i
\leftrightarrow \partial_i (\sim \bar{z}_i), q_i \leftrightarrow z_i$;
Poisson bracket in phase space $\{ p_i, q_j \} \leftrightarrow$ N.C.
$[\bar{z}_i, * z_j]$ on brane world sheet; C.M. module $p_i, q_j$ becomes
the central position of N.C. solitons on the brane. The finit shift
along two N.C. direction $(\frac{1}{n}, \frac{\tau}{n})$ causes the two
$U(n)$ matrices of $W_i$ wrapping two cycles around the compactified
sub-brane becomes noncommutative, so the target become a twisted
$su_n({\cal T})$ bundle section. Then the sources by the piercing
transverse strings, "impurities" further causes a $U(n)$ symmetry
breaking moment map, which reduces the N.C. $su_n({\cal T})$ cotangent
section field into the N.C. $L_{\rm CM}$. The modules of brane is
indentified to the N.C. soliton solution $q_i(t)$. These explain the
mysteries, why the SW curve corresponds to the spectral curve of C.M.
equation.

The Hamiltonian flows will determine the evolution of the brane $\Sigma$
in the ALF manifold $Q$ as $\theta(U_z + V_t + W) = 0$ \cite{K, KWZ},
where the period matrix of the Riemann $\theta$ function is the periods of
$B$-cycles of $\Sigma$; $U$, $V$ are some constant vectors \cite{K, KWZ,
GP} in the Jacobian of $\Sigma$, determined by the Abelian map of
meromorphic forms on $\Sigma$ and $W$ is the Riemann vector. Then the
explicit soliton solution $z_i(t)$ \cite{GP} will give the evolution of
the brane $\varphi^i_j(z(t))$ (\ref{varphi_ij}) in space time. We have
consider only the second order interaction as GMS \cite{GHS}. That is the
$V$ is a second jet, corresponds to the second order C.M. Hamiltonian
${\rm tr}|L^2|$, so the $z_i(t)$ is determined by the C.M. equation. The
exact Hamiltonian form should be determined by the action of brane.

The hyperquotient manifold $Q$ is obtained by the moment map, with the
moments (level) $\mu_i \sim$ the distance of $6$-branes $\hat{r}_i -
\hat{r}_{i+1}$, it is $A_{n - 1}$ Weyl invariant. We conject \cite{DM}
there exist an ALF orbifold such that $\mu_i$ can be identified with the
moments (FI source of D-flat equation) of C.M. That is it is reasonable to
conject the Kronheimer like relation for the moduli space of a ALF
orbifold with the moduli space of N.C. multisoliton ${\cal T}^{\otimes n}
/ {S_n}$.

The N.C. elliptic algebra geometry will faciliate to investigate
explicitly this correspondence and other properties of the ALF orbifold
and the K3 $Q(k, z)$ manifold.

\section{Quantum Hall effect on torus, Laughlin wave function of
quasiparticle, Bethe ansatz of elliptic Gaudin}

\indent

The QHE on ${\bf R}^2$ is described \cite{Susskind, Polychronakos} by
matrix (N.C.) Chern-Simons theory. The area preserveing gauge
transformation for the imcompressible electron fluid without source is
generated by the Gaussian constrain $[x_1, x_2] = \frac{i B}{2 \pi \rho_0}
\equiv i \theta B = \frac{i}{\nu}$, here $\rho_0$ is the density of the
carrier and $\nu$ the filling fraction.

After adding a quasiparticle source , the Gaussian constrain for finite
matrix Chern-Simons \cite{Polychronakos} turns to be
\begin{equation}
\label{CMeq}
B[x_1, x_2] = \frac{i}{\nu}(1 - n | v \rangle \langle v |),
\end{equation}
here the source is the boundary state $| v \rangle = \frac{1}{\sqrt{n}}(1,
1, \cdots, 1)$ or equivalently a Wilson line \cite{MP}. This (\ref{CMeq})
is the moment map for $su(n) \rightarrow u(n) / u(n - 1) \otimes U(1)$
\cite{GN}. Let $X_1 \equiv Q$ be diagonal, then $- B X_2 \equiv P$
becomes the rational C.M. Lax.

Hellerman and Raamsdonk \cite{HR} show the one-to-one correspondence in
symmetry pattern of Laughlin wave functions and the minimal basis of
matrix CS theory. Later Karabali and Sakita \cite{KS} find that, after
seperating the total antisymmetrical Vandermonde determinent representing
vacuum state, the matrix CS eigenfunction with given symmetry pattern is
identified to corresponding Jack polynomial representing the CM
eigenfunction.

In case of QHE on torus, the global shift $U$ should be considered as in
case of cylinder by Polychronakos \cite{Polychronakos2}. Now, both
the global shift $U_1, U_2$ and the local shift $X_1, X_2$ becomes N.C.
The Noncommutativity of $U_1, U_2$ implies the global $su(n)$ structure as
in section 3. While the holomorphic and antiholomorphic combination of
local shifts becomes respectively the connection form $\partial + A$ and
hermitian $n \times n$ matrix $\varphi$. Thus the constrain with a
quasiparticle source turns to be (\ref{mapeq}) which is the moment map
equation of the elliptic C.M. Furthermore, it is gauge (\ref{gaugem})
equivalent to the elliptic Gaudin with single marked point $u=0$. So, we
could find the quantum wave function explicitly by using the plentiful
known results of elliptic Gaudin-Calogero theory.

Etingof and Kirrilov \cite{EKi, EK} has shown that if the KZB equation of
the conformal block of twisted bundle on torus is restricted to the one
dimensional zero weight space of $S^{ln}C^n$, then it will be degenerate
into the elliptic C.M. equaiton. Subsequently, Felder and Varchenko find
the eigenfunctions as the integrand of the integral representations of the
solutions of KZB equation \cite{FV3}; and later \cite{FV4} as the Bethe
ansatz for the elliptic Gaudin, as the following theorem given by them.

{\bf Theorem}. {\it For an irreducible highest weight module with highest
weight $\Lambda = \sum_j m_j \alpha_j$ and highest weight vector
$v_\Lambda$, Set $m = \sum_j m_j$ and let the "color function" $c$: $\{ 1,
\cdots, m \} \rightarrow \{ 1, \cdots, n \}$ be the unique non-decreasing
function such that $c^{-1} \{ j \}$ has $m_j$ elements, for all $j = i,
\cdots, n$. Then the function parameterized by $t \in {\bf C}^m$
\begin{equation}
\label{Bethe-vector}
\psi (t, z) = e^{2 \pi i \zeta(z)} \sum_{\alpha \in S_n} w_{\sigma, c}(t,
\alpha_1(z), \cdots, \alpha_r(z)) f_{c(\sigma(1))} \cdots f_{c(\sigma(n))}
v_\Lambda
\end{equation}
is an eigenfunction of $H_{\rm Sug}$ (\ref{Sugawara}), if the parameters
$t_1, \cdots, t_n$ are a solution of the set of $n$ equations ("Bethe
ansatz equations")
\begin{equation}
\label{Betheeq}
\left ( \sum_{l : l \neq j} \frac{\theta^\prime(t_j - t_l)}{\theta(t_j -
t_l)}\alpha_{c(l)} - \frac{\theta^\prime(t_j)}{\theta(t_j)}\Lambda + 2 \pi
i \zeta, \alpha_{c(j)} \right ) = 0, \quad, j = 1, \cdots, n.
\end{equation}
The corresponding eigenvalue $\epsilon$ is
\begin{eqnarray}
\label{eigenvalue}
\epsilon & = & 4 \pi^2 (\zeta, \zeta) - 4 \pi i \frac{\partial}{\partial
\tau} S(t_1, \cdots, t_n, \tau),\nonumber\\
S(t_1, \cdots, t_m, \tau) & = & \sum_{i < j} (\alpha_{c(i)}, \alpha_{c(j)}
\ln \theta(t_i - t_j) - \sum_i (\Lambda, \alpha_{c(i)}) \ln \theta(t_i).
\end{eqnarray}}
Here, the "weight function" in (\ref{Bethe-vector}) $w_{\sigma, c}(t, z)
\equiv w(t_{\sigma(1)}, \cdots, t_{\sigma(n)}, z_{c(\sigma(1))}, \cdots,
z_{c(\sigma(n))})$ and
$$
w(t, z) = \prod_{j = 1}^m \frac{\theta(z_1 + \cdots + z_j - t_j + t_{j
+ 1})}{\theta(z_1 + \cdots + z_j) \theta(t_j - t_{j + 1})},
$$
and $\zeta_i$, $(i = 1, \cdots, n)$ are the twisting parameters
$$
(\zeta, \zeta) = \sum_i \zeta_i \zeta_i, \quad (\zeta, z) = \sum_i \zeta_i
z_i, \quad (\zeta, \alpha) = \zeta_j - \zeta_{j + 1}.
$$
The Sugawara Hamiltonian is
\begin{equation}
\label{Sugawara}
H_{\rm Sug} = - \sum_{i = 1}^n \partial_{z_i}^2 + \sum_{\alpha \in \Delta}
\wp (\alpha(z)) e_\alpha e_{- \alpha} + {\rm constant}.
\end{equation}
where $\Delta$ is the set of roots of $su(n)$, $e_\alpha \in sl(n)$ is the
corresponding root vector. For the primary root $\alpha_i$ $(i = 1,
\cdots, n - 1)$, $e_i \equiv e_{\alpha_i}$, $f_\alpha \equiv e_{-
\alpha}$, ($\alpha \in \Delta_+$), $f_i \equiv e_{- \alpha_i}$.

For given color function $c$, the Hermite-Bethe variety $HB(c)$ is defined
in \cite{FV2} by eliminating $\zeta$ from the BA equation (\ref{Betheeq}).
They show also that if $t_j$ is replaced by $t_j + l + m \tau$ $(l, m \in
Z)$, the $\zeta$ is shifted by $- m \alpha_{c(j)}$, and these replacements
do not change the eigenfunction $\psi$. Therefore there is a map $\zeta$:
$HB(c) \rightarrow h^* / Q$ ($Q$: root lattice of $A_{n - 1}$)
mapping $t$ to $\zeta$, and $HB(c)$ parameterizes eigenfunctions $\psi$
such that
\begin{equation}
\psi(z + \omega) = e^{2 \pi i \zeta(\omega)} \psi(z),
\end{equation}
here $\omega \in p^\vee$, $p^\vee = \{\omega \in h| \alpha(\omega)
\in Z \}$. Thus $\zeta$ determines the quasiperiodic twisting angle
$\theta$ of phase shift of $\psi(z)$ under a real periods of $z$ on torus.
It is wellknown that for the 8-vertex and $Z_n \times Z_n$ Belavin model,
the twisting angle $\theta$ between different sectors on different
"vacuum" of Bethe vector equals $2 \pi i z(\omega) \eta \equiv 2 \pi i
\lambda(\omega)$, $\eta$: crossing parameter, $\lambda = (\lambda_1,
\cdots, \lambda_n) \in h^*$, dynamical "heights" of different
"vacuum". In the trignometric limit $(\tau \rightarrow i \infty)$, this
twisting gives the exponents of the boudary matrix factor \cite{Sklyaninq}
of monodromy matrix for the twisted period boundary condition. This is
caused by the boundary term in the Hamiltonian. In Gaudin limit it
contributes a linear term $\zeta_i \delta_{i j}$ in $L_{i j}$ \cite{S},
and a shift $\zeta(\alpha_j)$ in Bethe ansatz \cite{ST} as in
(\ref{Betheeq}). This also explains the contributions $(\zeta, \zeta)$ for
the eigenvalue (\ref{eigenvalue}) of $H_{\rm Sug}$ (\ref{Sugawara}).

The Bethe ansatz solution (\ref{Bethe-vector}, \ref{Betheeq},
\ref{Sugawara}) is valid for generic $\zeta$. Now we turn to its
specialization in QHE. The papers \cite{EKi, FW, FV3} find the solution of
C.M. equation from asymptotic limit of the solution of KZB equation. Here
the crucial points is the "zero weight" and the vanishing condition
\cite{FW, FV3}. The Weyl invariance and vanishing condition is the
condition for solution of KZB equation, devided by the Weyl-Kac
denominator to be a conformal block of the $su_n({\cal T})$ WZW (i.e.
twisted $\hat{su}_n({\cal T})$ bundle), it requires \cite{FV3} $\zeta$ to
be dominant weight of $su(n)$. These conformal block gives the Bethe
ansatz solution \cite{FFR} of Gaudin model since the Gaudin Hamiltonian
(include e.g. the quadratic $H_{\rm Sug}$) spans the center of
$\hat{su}_n({\cal T})$ algebra at criticle level, this center is Poisson
isomorphic with the classical W algebra (include e.g. the Virasoro $H_{\rm
Sug}$). Their common eigenfunction is given by the Bethe ansatz.

Now we show that to be consistant with the Weyl invariant of the
$L$-matrix and moment map, the $\zeta$ vector should be $\zeta(1, \cdots,
1)$. In the algebra-geometric formulation \cite{K} the eigenvector $\psi$
(\ref{Bethe-vector}) corresponds to the Baker Akhiezer function $\psi(t,
z)$ of the Schr\"odinger equation
\begin{equation}
\label{S-eq}
(\partial_t - \partial_{z_i}^2 + 2 \sum_{i = 1}^n \wp(z - z_i(t))\psi = 0.
\end{equation}
The solution is expressed in terms of $n$ linear independent double-Bloch
solution with simple poles $z = z_i(t)$, and
\begin{equation}
\psi = \sum_{i = 1}^n c_i(t, k, u) w(z, z_i(t)) e^{kz + k^2 t},
\end{equation}
here $u$ is the spectral parameter (of $\Sigma_0$) and $k$ the conjugate
eigenvalue parameter in the spectral curve $\Sigma(k, u) \sim \Gamma(k,
u)$, $z = z_i(t), (i = 1, \cdots, n)$ is the $n$ independent solutions of
classical C.M. equation. In fact the vanishing of the double poles in
(\ref{S-eq}) gives
\begin{equation}
(L_{\rm CM}(t, u) - k I)c = 0, \quad c = (c_1, \cdots, c_n).
\end{equation}
Here, the $L_{\rm CM}$ is identical to (\ref{L_CM}), as $\dot{z}_i = 2
p_i$. Then as described in the last subsection, the $L_{\rm CM}(u)$ at $u
= 0$ has the form
\begin{equation}
\zeta(1 - n |v \rangle \langle v |)\frac{1}{u} + O(1),
\end{equation}
which is the consequence of the moment map condition (\ref{mapeq}), i.e.
the {\it quasipaticle source} of the {\it Gauss constrain} for QHE.

Finnally let us match with the {\it Laughlin wave functions}. On the zero
weight space the KZB connection for heat equation becomes \cite{FW} the
quadratic Gaudin Hamiltonian $H_{\rm Sug}$, and on the module $S^{ln}C^n,
i.e. \Lambda = l \sum_{j - 1}^{n - 1} (n - j) \alpha_j$, the zero weight
space is one dimensional \cite{EK}, and $H_{\rm Sug}$ reduces \cite{EKi,
FW} to the C.M. Hamiltonian (\ref{CM-Hamiltonian}) with coupling constant
$g = \zeta^2 \cong l^2$. Then it is easy to see that the $(l + 1)$th power
of the total antisymmetrical Kac-Weyl denominator is a horizontal section
for the KZB equation, i.e. the vacuum vector (ground state of QHE) of the
quantum C.M. equation. The quantum coupling constant of C.M. is $l(l+1)$,
and $\frac{1}{(l+1)}$ is the filling fraction in QHE.

At the criticle level the B.A. eigenfunction (\ref{Bethe-vector},
\ref{Betheeq}, \ref{Sugawara}) turns to be that of C.M. $H$
(\ref{CM-Hamiltonian}) or $H_{\rm Sug}$ (\ref{Sugawara}) respectively for
the $S^{ln}C^n$ or the adjoint representation. The zero weight space of
the later, the adjoint representation is $n$ dimensional.

The corresponding color function $c$, gives the partition number of
oscillators (creation operators ${\rm tr}((A^\dagger)^i)$ of \cite{HR}),
so determines the symmetrical pattern of the "elliptical" generalized Jack
polynomial for the Laughlin wave function on torus, as has been
conjectured by Hellerman and Raamsdonk \cite{HR} and explicitely given by
Karabali and Sakita \cite{KaSa} for the N.C. $R^2$ case.

\section{Discussion}

\indent

The aim of this paper is to show that once the N.C. integrable model has
been found for N.C. torus (and sphere), then the various algebraic
analytic and algebraic geometrical methods can be used to explicitely
investigate the relavent physics for string-brane etc. Proviously in a
talk given by B. Y. Hou \cite{HP}, we have given the multi-soliton
solutions on N.C. torus in connection with the elliptic algebra for
integrable models. The detail and expanded version of this talk has been
put in the web as \cite{HPSY}, which has never been submitted elsewhere.
In the section 3 there in, the algebra ${\cal A}_n$ and the Hilbert space
${\cal H}_n$, in reality is relevant only for the vortex free (zero moment
map) trivial bundle of the $su_n({\cal T})$ symmetry algebra. While the
elliptic Gaudin and C.M. are $su(n)$ bundle constrained by moment map with
source. We included this as the section 3 and 4 of this paper.

Now this paper further display explicitely how these algebras work for
brane and QHE. We have given the ${\cal H}_n$ and ${\cal A}_n$ of the LLL
(lowest Landau level) on the $S^4$ \cite{CHH} in the vortex free case.
We're working of the LLL Laughlin wave functions in case with
quasiparticles source both for the fuzzy K\"ahler $S^2$ and for the
hyperK\"ahler $S^4$.

Meanwhile we have a preliminary result of the projection operators of
multisolitons on the $Z_k$ orbifolding N.C. torus \cite{HSYa}. The result
of this paper for brane, Laughlin wave function etc. will be easily
generalized into the descrete orbifold and orientfold. Then, by pairing
these moment map of impurity sources to the corresponding moment map
of gravity centers of ALF, one may investigate various behavior including
the dual properties of the brane configuration.

But to investigate the full Narain duality, we must consider the case of
the more generic $\theta \neq \frac{Z}{n}$ case. We have argued in the
second part of our unsubmitted paper \cite{HPSY}, that the generic
$\theta$ will corresponding to the ratio of modulus of the two $SL(2 Z)$
in Narain's $SL(2 Z) \otimes SL(2 Z)$ and then to the crossing parameter
$\eta$ of $Z_n \otimes Z_n$ Belavin and RSOS and to the coupling constant
of the elliptic Ruijseenaar-Schneider model. These should be checked more
carefully in connection with the brane theory.

At last we would like tentatively remind that in the Wakimoto
construction for conformal field, the zero mode is related to the N.C.
part $A_1$ of Witten \cite{W}, i.e. the boundary conformal field in
Seiberg-Witten limit. It is interesting to find the relation of the boson
oscillators which realizes the $A_0$ part or the string brane
interactions.


\begin{thebibliography}{[99]}

\bibitem{H} J. A. Harvey, {\it Komaba Lectures on Noncommutative
Solitons and D-Branes} {\bf hep-th/0102076}.

\bibitem{DN} M. R. Douglas, N. A. Nekrasov, {\it Noncommutative Filed
Theory}, {\bf hep-th/0106048}.

\bibitem{GMS} R. Gopakumar, S. Minwalla, A. Strominger, {\bf
hep-th/0003160}, JHEP 0005:020(2000).

\bibitem{W} E. Witten, {\it Noncommutative Tachyons and String Field
Theory}, {\bf hep-th/0006071}.

\bibitem{HKL} J. A. Harvey, P. Kraus. F. Larsen, {\bf hep-th/0010060},
JHEP 0012:024(2000).

\bibitem{BKMT} I. Bars, H. Kajiura, Y. Matsuo, T. Takayanagi, Phys. Rev. D
{\bf 63}(2001)086001, {\bf hep-th/0010101}.

\bibitem{SS} E. M. Sahraoui, E. H. Saidi, {\it Solitons on Compact And
Noncompact Spaces in Large Noncommutativiy}, {\bf hep-th/0012259}.

\bibitem{MM} E. J. Martinec, G. Moore, {\it Noncommutative Solitons on
Orbifolds}, {\bf hep-th/0101199}.

\bibitem{B} F. P. Boca, Comm. Math. Phys. {\bf 202}(1999), 325.

\bibitem{GHS} R. Gopakumar, M. Headrick, M. Spradlin, {\it On
Noncommutative Multi-solitons}, {\bf hep-th/0103256}.

\bibitem{TKMS} T. Krajewski, M. Schnabl, {\it Exact Solitons on
Noncommutative Tori}, {\bf hep-th/0104090}.

\bibitem{KMT} H. Kajiura, Y. Matsuo, T. Takayanagi, {\bf hep-th/0104143},
JHEP 0106:041(2001).

\bibitem{Bars} I. Bars, {\it Nonpertubative Effects of Extreme
Localization in Noncommutative Geometry}, {\bf hep-th/0109132}.

\bibitem{CDS} A. Connes, M. R. Douglas, A. Schwarz, {\bf hep-th/9711162},
JHEP 9802:003(1998).

\bibitem{SW} N. Seiberg, E. Witten, {\bf hep-th/9908142}, JHEP
9909(1999)032.

\bibitem{R} M. Rieffel, Pacific J. Math. {\bf 93} (1981) 415.

\bibitem{KS} A. Konechny, A. Shwarz, {\it Introduction to matrix theory
and noncommutative geometry}. {\bf hep-th/0012145} and {\bf
hep-th/0107251}.

\bibitem{DKL} L. Dabrowski, T. Krajewski, G. Landi, Int. J. Mod. Phys.
B{\bf 14} (2000) 2367.

\bibitem{BCZ} H. Bacry, A. Grossman, J. Zak, Phys. Rev. {\bf B12}(1975)
1118.

\bibitem{Zak} J. Zak, in Solid State Physics, edited by H. Ehrenreich, F.
Seitz and D. Turnbull (Academic, New York, 1972), Vol. 27.

\bibitem{FV4} G. Felder, A. Varchenko, {\it Three formulas for
eigenfunctions of integrable Schr\"odinger operators}, {\bf
hep-th/9511120}.

\bibitem{Perelomor} A. M. Perelomov, {\it Generalized Coherent State and
Their Applications}, Springer-Verlag, Berlin, 1986.

\bibitem{Tata} D. Mumford, Tata lecture on Theta I. Basel-Boston:
Birkha\"user 1982.

\bibitem{AMNS} J. Ambjorn, Y.M. Makeenko, J. Nishimura, R. J. Szabo, JHEP
0005 (2000) 023, {\bf hep-th/0004147}.

\bibitem{BA} D. Bigatti, {\it Gauge Theory on the Fuzzy Sphere}, {\bf
hep-th/0109018},

\bibitem{CFH} K. Chen, H. Fan, B. Y. Hou, K. J. Shi, W. L. Yang, R. H.
Hong, Prog. Theor. Phys. Supp. 135(1999)149.

\bibitem{Sklynin} E. K. Sklyanin, Funct. Annal. Appl. {\bf 16}(1982)263,
ibid. {\bf 17}(1983)273.

\bibitem{ST} E. K. Sklyanin, T. Takebe, {\it Separation of Variables in
the Elliptic Gaudin Model}, {\bf solv-int/9807008}.

\bibitem{Kac} V. Kac, {\it Infinite Dimensional Lie Algebras}, Cambridge
University Press, 1990.

\bibitem{HM} J. C. Hurtubise, E. Markman, Comm. Math. Phys. {\bf
233} (2001) 533, {\bf math.AG/9912161}.

\bibitem{ST1} E. K. Sklyanin, T. Takebe, Phys. Lett. A{\bf 219}(1996)217.

\bibitem{ER} B. Enriquez, V. Rubetsov, {\it Hitchin Systems, Higher Gaudin
Operators and $r$-matrices}, {\bf alg-geom/9503010}.

\bibitem{KT1} G. Kuroki, T. Takebe, {\it Bosonization and integral
representation of solutions of the Kniznik-Zamolodchikov equation}. {\bf
math.QA/9809157}.

\bibitem{HPSY} B. Y. Hou, D. T. Peng, K. J. Shi, R. H. Yue, {\it Solitons
on Noncommutative Torus as Elliptic Algebra and Elliptic Models}. {\bf
hep-th/0110122}.

\bibitem{FFR} B. Feigin, E. Frenkel, N. Reshetikhen, Comm. Math. Phys.
{\bf 166} (1994) 27.

\bibitem{FW} G. Felder, C. Wieczerkowski, Comm. Math. Phys. {\bf 176}
(1996) 133.

\bibitem{KT} G. Kuroki, T. Takebe, Comm. Math. Phys. {\bf 190} (1997) 1.

\bibitem{BCS} A. J. Bordner, E. Corrigan, R. Sasaki, Prog. Theor. Phys.
{\bf 102} (1999) 499.

\bibitem{BFV} V. M. Buchstaber, G. Felder, A. V. Veselov, Duke Math. J.
{\bf 76} (1994) 885.

\bibitem{KSe} A. Kapustin, S. Sethi, Adv. Theor. Math. Phys. {\bf 2}
(1998) 571. {\bf hep-th/9804027}.

\bibitem{DM} M. R. Douglas, G. Moore, {\it D-branes, Quivers, and ALE
Instantons}, {\bf hep-th/9603167}.

\bibitem{BBST} B. A. Bernevig, J. Brodie, L. Susskind, N. Toumbas, {\it
How Bob Laughlin Tamed the Giant Graviton from Taub-NUT space}, JHEP 0102
(2001) 003, {\bf hep-th/0010105}.

\bibitem{Susskind} L. Susskind, {\it The Quantum Hall Fluid and
Non-Commutative Chern Simons Theory}, {\bf hep-th/0101029}.

\bibitem{Polychronakos} A. P. Polychronakos, {\it Quantum Hall states as
matrix Chern Simons theory}, JHEP 0104 (2001) 011, {\bf hep-th/0103013}.

\bibitem{GKK} A. Gorsky, I. I. Kogan, C. Korthels-Altes, {\it Dualities in
Quantum Hall System and Noncommutative Chern-Simons Theory}, JHEP 0201
(2002) 002, {\bf hep-th/0111013}.

\bibitem{MP} B. Morariu, A. P. Polychronakos, {\it Finite Noncommutative
Chern-Simons with a Wilson Line and the Quantum Hall Effect},

\bibitem{Krichver} I. M. Krichver, Fun. Ann. App. 14 (1980) 282.

\bibitem{DP1} E. D'Hoker, D. H. Phong, Asian J. Math. 2 (1998) 655, {\bf
hep-th/9808156}.

\bibitem{DP2} E. D'Hoker, D. H. Phong, Lectures on supersymmetric
Yang-Mills Theory and Integrable systems, {\bf hep-th/9912271}.

\bibitem{D} D. Diaconesqn, {\it D-Branes, Monopoles and Nahm equations},
{\bf hep-th/9608163}.

\bibitem{HW} A. Hanany, E. Witten, Nucl. Phys. B {\bf 492} (1997) 652.
{\bf hep-th/9611230}.

\bibitem{Witten} E. Witten, Nucl. Phys. B {\bf 500} (1997) 3. {\bf
hep-th/9703166}.

\bibitem{GR} G. W. Gibbons, P. Rychenkova, {\it Hyper K\"ahler quotient
construction of BPS Monopole Moduli spaces}, {\bf hep-th/9608085}.

\bibitem{DW} R. Donagi, E. Witten, Nucl. Phys. B {\bf 460} (1996) 299,
{\bf hep-th/9510101}.

\bibitem{MMM} A. Marshakov, M. Martellini, A. Morzov, Phys. Lett. A {\bf
418} (1998) 294, {\bf hep-th/9706050}.

\bibitem{K} I. Krichever, {\it Elliptic solutions to difference non-linear
equations and nested Bethe ansatz equations}, {\bf solv-int/9804016}.

\bibitem{KP} I. M. Krichver, D. H. Phong, {\it Sympletic forms in the
theory of solitons}, {\bf hep-th/9708170}.

\bibitem{KWZ} I. Krichever, P. Wiegmann, A. Zabrodin, Comm. Math. Phys.
{\bf 193} (1998) 373.

\bibitem{GP} L. Gavrilov, A. M. Perelomov, {\it On the explicit
solutions of the elliptic Calogera-Moser systems}, {\bf solv-int/9905051}.

\bibitem{GN} A. Gorsky, N. Nekrasov, Nucl. Phys. B {\bf 414} (1994) 213,
{\bf hep-th/9401021}.

\bibitem{HR} S. Hellerman, M. V. Raamsdonk, JHEP 0110 (2001 039, {\bf
hep-th/0103179}.

\bibitem{Polychronakos2} A. P. Polychronakos, {\it Quantum Hall states on
the cylinder as unitary matrix Chern-Simons theory}, {\bf hep-th/0106011}.

\bibitem{EKi} P. Etingof, A. Kirillov Jr, Duke. math. J. {\bf 74} (1994)
585, {\bf hep-th/9310083}.

\bibitem{EK} P. I. Etingof and A. A. Kirillov, Jr., Math. Res. Lett. {\bf
1} (1994), 179. {\bf hep-th/9403168}.

\bibitem{FV3} G. Felder, A. Varchenko, Int. Math. Res. notices {\bf
5} (1995) 221, {\bf hep-th/9502165}.

\bibitem{FV2} G. Felder, A. Varchenko, Nucl. Phys. B {\bf 480}(1996) 485.

\bibitem{Sklyaninq} E. K. Sklyanin, {\it Quantum Inverse Scattering
method. Selected topics}. {\bf hep-th/9211111}.

\bibitem{S} E. K. Sklyanin, J. Sov. Math. {\bf 47} (1989) 2473.

\bibitem{KaSa} D. Karabali, B. Sakita {\it Orthogonal basis for the energy
eigenfunctions of the Chern-Simons matrix model}. {\bf hep-th/0107168}.

\bibitem{HP} B. Y. Hou, D. T. Peng, {\it Elliptic algebra and Integrable
Models for solitons on Noncommutative Torus ${\cal T}$}. Talk given by B.
Y. Hou at the Joint APCTP-Nankai Symposium, Tianjin, China, Oct. 2001. To
apear in the proceedings, to be published by Int. J. Mod. Phys. B. {\bf
hep-th/0111094}.

\bibitem{CHH} Y. X. Chen, B. Y. Hou, B. Y. Hou, {\it Noncommutative
geometry of 4-dimensional quantum Hall droplet}. {\bf hep-th/0203095}.

\bibitem{HSYa} B. Y. Hou, K. J. Shi, Z. Y. Yang, {\it Solitons on
Noncommutative Orbifold $T^2 / Z_k$}, {\bf hep-th/0204102}.

\end{thebibliography}
\end{document}